\documentclass[sigconf]{acmart}
\usepackage{listings}
\lstdefinelanguage{JavaScript}{
  keywords={break, case, catch, continue, debugger, default, delete, do, else, false, finally, for, function, if, in, instanceof, new, null, return, switch, this, throw, true, try, typeof, var, void, while, with},
  sensitive=true,
  comment=[l]{//},
  morecomment=[s]{/*}{*/},
  morestring=[b]", morestring=[b]',
  morestring=[b]`,
}

\AtBeginDocument{%
  }

\setcopyright{acmlicensed}
\copyrightyear{2025}
\acmYear{2025}
\acmDOI{XXXXXXX.XXXXXXX}
\acmConference[WiseML 2025]{ACM Workshop on Wireless Security and Machine Learning}{June 30-- July 03,
  2025}{Arlington, VA}
\acmISBN{978-1-4503-XXXX-X/2018/06}

\begin{document}

\title{Leveraging LLM to Strengthen ML-Based Cross-Site Scripting Detection\\}


\author{Dennis Miczek, Divyesh Gabbireddy, Suman Saha}
\affiliation{%
  \institution{Pennsylvania State University}
  \city{University Park}
  \state{PA}
  \country{USA}
}
\email{{dkm6080, dmg6433, szs339}@psu.edu}

% \author{Anonymous4}
% \affiliation{%
%   \institution{dept. name of organization (of Aff.)}
%   \city{city}
%   \state{State}
%   \country{Country}
% }
% \email{email@address}

% \author{Anonymous5}
% \affiliation{%
%   \institution{dept. name of organization (of Aff.)}
%   \city{city}
%   \state{State}
%   \country{Country}
% }
% \email{email@address}

\renewcommand{\shortauthors}{Surname et al.}

\begin{abstract}
According to the Open Web Application Security Project (OWASP), Cross-Site Scripting (XSS) is a critical security vulnerability. Despite decades of research, XSS remains among the top 10 security vulnerabilities. Researchers have proposed various techniques to protect systems from XSS attacks, with machine learning (ML) being one of the most widely used methods. An ML model is trained on a dataset to identify potential XSS threats, making its effectiveness highly dependent on the size and diversity of the training data. 

A variation of XSS is obfuscated XSS, where attackers apply obfuscation techniques to alter the code's structure, making it challenging for security systems to detect its malicious intent. Our study's random forest model trained on traditional (non-obfuscated) XSS data achieved 99.8\% accuracy. However, when tested against obfuscated XSS samples, accuracy dropped to 81.9\%, underscoring the importance of training ML models with obfuscated data to improve their effectiveness in detecting XSS attacks. A significant challenge is to generate highly complex obfuscated code despite the availability of several public tools. These tools can only produce obfuscation up to certain levels of complexity.  

In our proposed system, we fine-tune a Large Language Model (LLM) to generate complex obfuscated XSS payloads automatically. By transforming original XSS samples into diverse obfuscated variants, we create challenging training data for ML model evaluation. Our approach achieved a 99.5\% accuracy rate with the obfuscated dataset. We also found that the obfuscated samples generated by the LLM were 28.1\% more complex than those created by other tools, significantly improving the model's ability to handle advanced XSS attacks and making it more effective for real-world application security.
\end{abstract}





\maketitle

\section{Introduction}

People spend a lot of their daily time online, using various mobile browsers or applications for work, banking, shopping, and social networking. These applications often create dynamic content that adapts based on user inputs, making it essential to validate any input before it is processed within the system. Unfortunately, many applications do not adequately sanitize user input, which exposes users to cybersecurity risks like command injection. Command injection involves an attacker embedding malicious code into a vulnerable program, which can lead to the execution of unintended commands or unauthorized access to sensitive data. According to the Open Web Application Security Project (OWASP), command injection ranked third in the top ten severe cyber security issues \cite{owasp2021}. Cross-site Scripting (XSS) is one of the variations of command injection. It occurs when an attacker injects malicious scripts, typically written in JavaScript, into web pages that other users view. These scripts run in the user's browser without their consent. The potential consequences of XSS are severe, including stealing user cookies to hijack sessions, accessing sensitive data, or even allowing attackers to take control of devices.

Despite years of research and the development of various protective measures \cite{wassermann2008static}\cite{tripp2009taj}\cite{vogt2007cross}\cite{jovanovic2006pixy} \cite{hallaraker2005detecting}\cite{kirda2006noxes}, XSS remains a serious concern. The continuous efforts in combating XSS are evident, but one challenge is the complexity of certain attack payloads, which can bypass conventional detection mechanisms. A particularly problematic form is obfuscated XSS code, where attackers deliberately modify the code to make it difficult to read and understand \cite{fang2018deepxss}. Techniques used for obfuscation include renaming variables with random strings, altering the control flow to obscure the execution order, inserting unrelated code blocks, and splitting strings to hide recognizable patterns. Such techniques make it easier for attackers to bypass security measures and execute harmful actions within vulnerable systems \cite{xu2012power}. 

One widely used technique for protecting against Cross-Site Scripting (XSS) is using machine learning or deep learning models trained to detect XSS code automatically \cite{fang2018deepxss}\cite{abaimov2019coddle}\cite{wang2014machine}\cite{kascheev2020detecting}\cite{lei2020xss}. The effectiveness of these models largely relies on the size and variety of the training dataset. In our experiment, we used a random forest learning approach to train a model on a dataset consisting of 19,359 samples. We observed an impressive accuracy rate of 99.8\% when tested with standard data. However, when we evaluated the model using an obfuscated dataset version, the accuracy dropped to 81.9\%. Most obfuscated XSS payloads that evaded protection were often long and very complex. However, we observed that the model failed to detect even simple changes in the payload. Even with simple methods, such as encoding a string in base64, can be effective against XSS detection in some cases. For example, a payload in listing \ref{lst:non-obs} was detected before obfuscation but evaded detection after obfuscation.

\begin{lstlisting}[language=JavaScript, caption={XSS payload in non-obfuscated and obfuscated versions}, label={lst:non-obs}]

// non-obfuscated code
<keygen autofocus onfocusin=alert(1)>

// Obfuscated code
<keygen autofocus onfocusin=eval(atob(
'ZmV0Y2goJ2h0dHBzOi8vZ29vZ2xlLmNvbS9s
b2c/Y29va2llPScgKyBkb2N1bWVudC5jb29
raWUpOw=='))>

\end{lstlisting}

\noindent This outcome underscores the crucial role of dataset diversity in building robust detection systems. The challenges in training machine learning models for XSS detection is sourcing actual XSS code, particularly obfuscated variants. As a result, machine learning researchers often rely on publicly available tools to generate obfuscated datasets for training and testing their models\cite{tellenbach2016detecting}. While these tools can produce obfuscated code effectively, their ability to create highly complex samples is limited, impacting the robustness of the resulting detection systems. Creating a comprehensive and complex obfuscated dataset is essential to developing resilient machine-learning models to identify advanced XSS attacks.

Our research addresses the significant challenge of constructing a robust machine-learning model for detecting XSS by leveraging Large Language Models (LLMs). LLMs are known for their extensive use in natural language processing and have proven to be highly effective for tasks involving code generation and analysis\cite{ibrahimzada2023automated}\cite{martin2008automatic}\cite{kang2023large}\cite{hossain2024deep}\cite{wen2024automatically}\cite{mathews2024llbezpeky}. We introduce a novel approach that fine-tunes an LLM to generate obfuscated versions of original XSS samples, creating a diverse and complex training dataset to enhance the robustness of machine learning models. Our method has demonstrated a high success rate, achieving 99.5\% accuracy when training and testing a model with the obfuscated dataset generated by LLM. Additionally, we found that the LLM-generated obfuscated samples were 28.1\% more complex than those produced by existing tools. This significant increase in complexity enhances the model's capability to identify sophisticated XSS attacks, making it a potent tool for improving real-world application security.

The rest of this paper is organized as follows: Section II details our research approach. Section III presents the results of our experiments, showcasing the effectiveness and improvements in model performance. Section IV discusses related work, focusing on studies that utilize LLMs for code generation, vulnerability identification, and system testing. Finally, we conclude our findings and suggest directions for future research in Section V.

\section{Approach}

This section discusses 1) the dataset, 
2) selecting machine-learning models,
3) the obfuscation techniques used to generate obfuscated code for both initial experiments and fine-tuning the LLM, and
4) generating Obfuscated XSS by LLM

\subsection{Dataset}
For this study, we compiled a robust dataset by aggregating multiple sources to cover a wide range of benign and malicious XSS payloads presented in Table \ref{tab:dataset}. After removing duplicates and applying strong filters on benign sources, we obtained a final dataset of 19,359 examples, with 12,038 labeled benign (62.18\%) and 7,321 as malicious (37.82\%). This dataset provides a well-rounded selection of XSS attack patterns and benign JavaScript code, covering typical and edge-case XSS vectors. A 62:38 benign-to-malicious ratio maintains a realistic balance crucial for training an effective detection model that minimizes false positives while accurately identifying malicious payloads.

\begin{table}[htbp]
\caption{The sources of benign and malicious payloads}
\label{tab:dataset}
\resizebox{\columnwidth}{!}{%
\begin{tabular}{lccc}
  \toprule
  \textbf{Dataset Source} & \textbf{Benign} & \textbf{XSS} & \textbf{Total} \\

  & \textbf{code} & \textbf{payloads} &  \\

  \midrule
  Kaggle$^{a}$ & 6,313 & 7,373 & 13,686 \\
  JS library source code$^{b}$ & 11,120 & -- & 11,120 \\
  XSS Cheat Sheet$^{c}$ & -- & 6,047 & 6,047 \\
  Materialize JS library$^{d}$ & 6,752 & -- & 6,752 \\
  \midrule
  \textbf{Total} & \textbf{24,185} & \textbf{13,420} & \textbf{37,605} \\
  \bottomrule
  \multicolumn{4}{l}{$^{a}$XSS dataset for Deep learning by Syed Saqlain Hussain Shah.} \\
  \multicolumn{4}{l}{$^{b}$https://github.com/twbs/bootstrap} \\
  \multicolumn{4}{l}{$^{c}$https://portswigger.net/web-security/cross-site-scripting/cheat-sheet} \\
  \multicolumn{4}{l}{$^{d}$https://github.com/Dogfalo/materialize} \\
\end{tabular}%
}
\end{table}

To prepare the dataset for training, we applied several preprocessing steps. First, we normalized the text by converting all payloads to lowercase, removing extra spaces, and stripping newline characters. Next, we eliminated redundant samples to ensure each payload was unique. We then split the cleaned dataset into a training set containing 80\% (15,487 samples) and a testing set containing 20\% (3,872 samples) to ensure a fair evaluation of our models. Afterward, we transformed the training and testing data separately into a structured numerical representation using a bag-of-words (BoW) approach, implemented with \texttt{CountVectorizer} and the token pattern \texttt{r"(?u)\textbackslash b\textbackslash w+\textbackslash b"}. This comprehensive preprocessing ensures data consistency and minimizes noise introduced by syntax variations, enabling practical model training and evaluation.

\subsection{Machine Learning Model Selection}
We selected four different machine-learning models for our study. 
1) \textit{Decision Tree} machine-learning model is a simple yet powerful model that makes decisions by splitting the dataset into branches based on feature values, ultimately leading to a decision at the leaf nodes. 
2) \textit{Support Vector Machine (SVM)} is a supervised learning model that finds the optimal hyperplane to separate data points into different classes. 
3) \textit{Logistic Regression} is a statistical model used for binary classification that predicts the probability of a sample belonging to one of two classes.
4) \textit{Random Forest} is an ensemble learning technique that builds multiple decision trees during training and combines their outputs for more robust predictions. This approach helps improve accuracy and reduces the risk of overfitting by averaging the results of various decision trees.

\subsection{Obfuscation Methods}

We applied multiple obfuscation techniques to generate JavaScript code samples to evaluate the robustness of our machine-learning models against obfuscated attacks. These included JavaScript obfuscation, where code transforms by modifying control flow, injecting irrelevant code, and renaming variables to hinder readability and detection. We also utilized \texttt{Base64} encoding, converting payloads into encoded strings decoded and executed dynamically via JavaScript's \texttt{eval(atob())}. Additionally, \texttt{URL} encoding was employed, encoding payloads into URL-safe formats and dynamically executing them using \texttt{eval(decodeURIComponent())}. Finally, we used a method that randomly splits payload strings and recombines them at runtime to obscure detection patterns. These obfuscation techniques effectively represent realistic adversarial strategies, challenging machine learning models to detect malicious JavaScript payloads despite obfuscation robustly.

\subsection{Generate Obfuscated XSS by LLM}

We chose CodeT5-small model for its unique advantages in code-related tasks, standing out from general-purpose language models such as GPT-4, which primarily focus on natural language. CodeT5-small is specifically pre-trained on code, enabling it to grasp intricate code structures, including syntax and logical flow. This specialization enhances its ability to generate sophisticated code obfuscations. Notably, its identifier-aware pre-training ensures that semantic integrity is preserved during code transformations—a crucial feature for producing obfuscated cross-site scripting (XSS) payloads that are structurally distinct but functionally consistent. Additionally, with its manageable size of 60 million parameters, CodeT5-small effectively balances performance and computational efficiency. This makes it ideal for fine-tuning and rapid experimentation on hardware with limited compute and memory capacity without compromising output quality or performance. 

\begin{figure}[htbp]
\centerline{\includegraphics[width=75mm,scale=0.5]{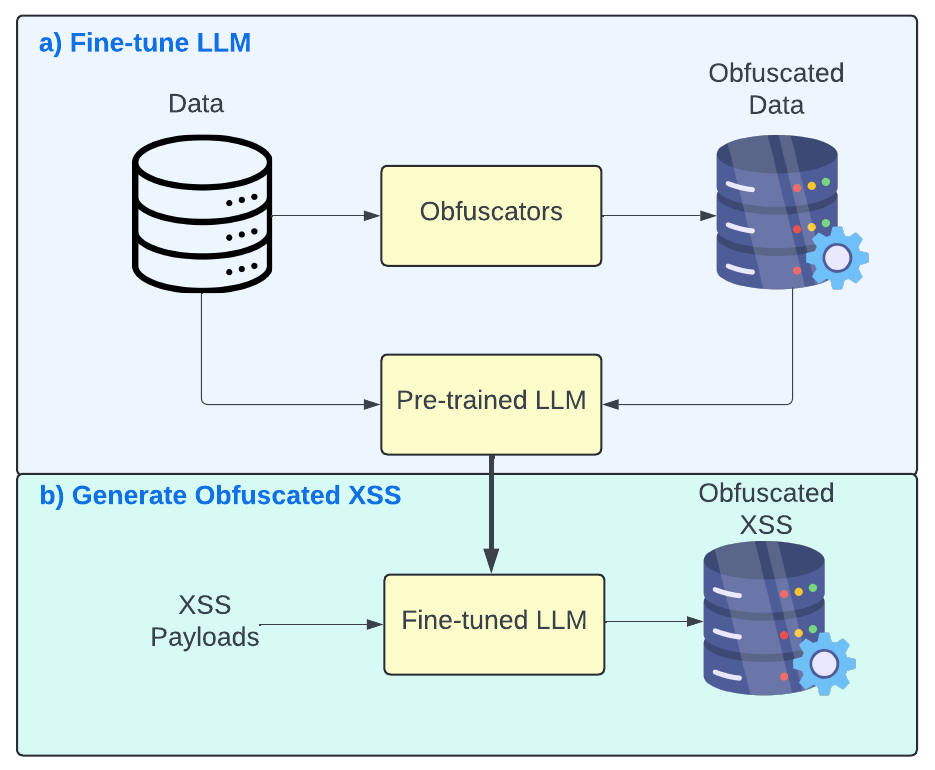}}
\caption{
a) The original data is obfuscated using obfuscation tools, and both the original and obfuscated versions are used to fine-tune a pre-trained LLM.
b) The fine-tuned LLM is then used to generate obfuscated XSS payloads from the original XSS payloads.}
\label{fig-approach}
\end{figure}

We developed a comprehensive methodology for fine-tuning to enhance the CodeT5-small model’s ability to generate diverse and challenging obfuscated XSS payloads, presented in Fig \ref{fig-approach}a. First, we prepared the training data by transforming the original XSS payloads into obfuscated versions using a combination of publicly available obfuscation tools and custom Python scripts. The obfuscation techniques applied included base64 encoding, URI encoding, and keyword splitting. These methods were selected to ensure that the resulting obfuscated payloads maintained their original functionality but were structurally different.

We initialized the CodeT5-small pre-trained model using the transformers library and fine-tuned it on a meticulously curated dataset containing original and obfuscated XSS samples. Key training parameters—including learning rates, batch sizes, and the number of epochs—were optimized to maintain high output quality while ensuring computational efficiency. This step ensured the model learned to alter the appearance of code while preserving its functionality. Performance was validated at regular intervals to confirm that the fine-tuned model generated syntactically correct and effective obfuscations.

After fine-tuning the CodeT5-small model, we employed it to generate obfuscated variants of XSS code from an initial set of base payloads, as depicted in Fig. \ref{fig-approach}b. The generation process involved using specially crafted prompts for the fine-tuned LLM to create obfuscated XSS payloads that preserved the original functionality while adding structural complexity. Specific tokens were used in the prompts to ensure the core behavior of the payloads remained unchanged. We varied the sampling strategies to produce diverse outputs and applied higher temperature parameters during generation. Temperature settings allowed for controlled randomness in the outputs, with higher temperatures generating more varied and less predictable obfuscated code. 
% The generated obfuscated code underwent a thorough manual review to confirm that it maintained semantic equivalence with the original payloads. 
The fine-tuned LLM introduced unique formatting and structural changes to the payloads, such as altering whitespace, inserting non-functional characters, or changing encoding methods. These variations created a diverse and challenging dataset, improving the efficacy of machine learning models when trained with both traditional and LLM-generated obfuscated datasets.

\section{Experiment}

All experiments in this study were performed on a system featuring an Intel(R) Xeon(R) CPU @ 2.20GHz, 12.7 GB of RAM, and a Tesla T4 GPU for enhanced processing power. The high-performance GPU played a crucial role in fine-tuning the LLM and generating obfuscated data, enabling efficient handling of extensive datasets and complex model training. This setup provided the computational capacity needed to  train the LLM on a diverse and large dataset while ensuring training times were kept manageable. We utilized Weights \& Biases \cite{wandb} to monitor and track the results of the experiments throughout the study.

\subsection{Evaluate Models without Obfuscated XSS}

\begin{figure}[h]
  \centering
  \includegraphics[width=\linewidth]{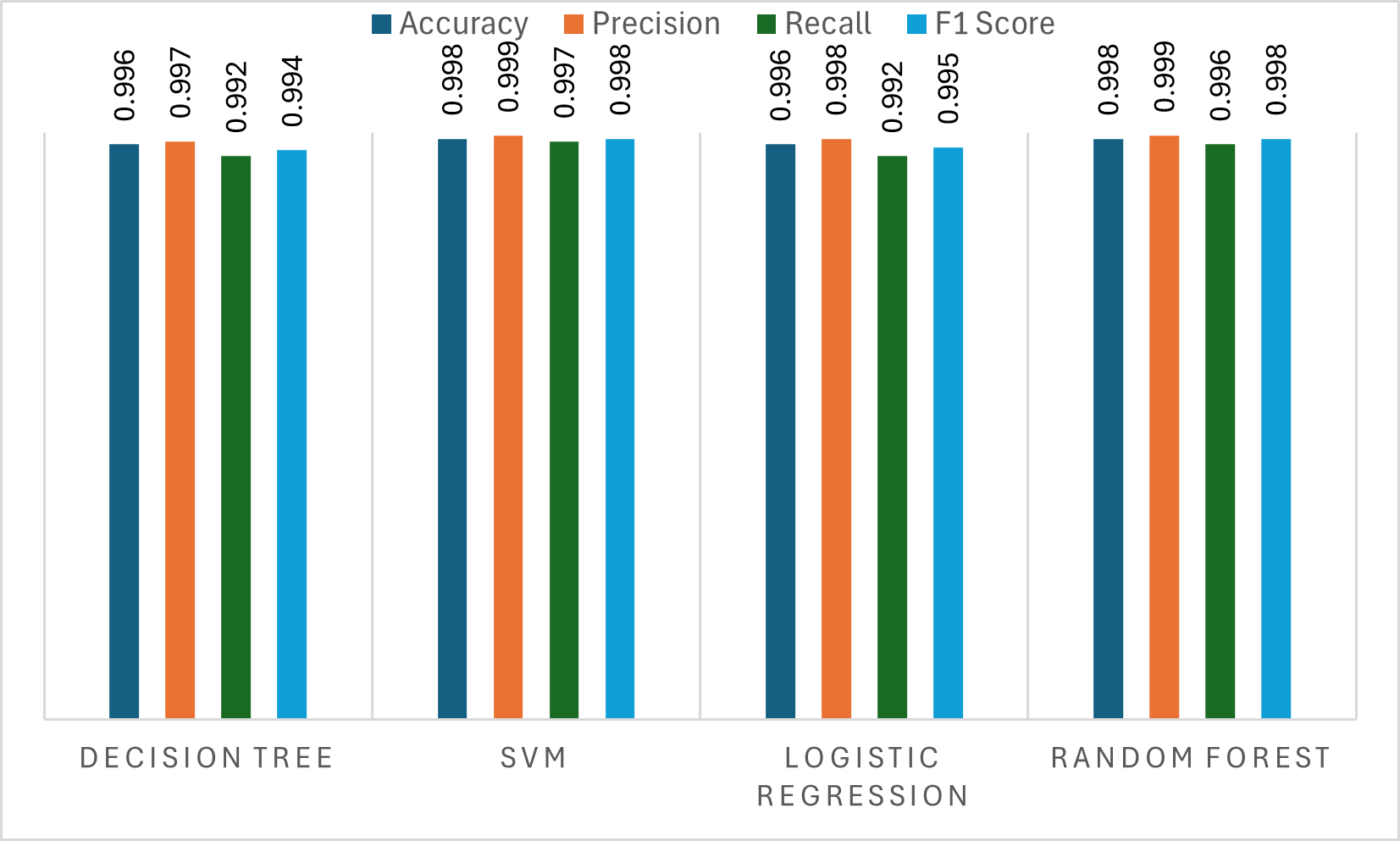}
  \caption{The performance of models when trained and tested without obfuscated XSS}
  \label{fig-exp-without-obfuscated}
\end{figure}

We chose four types of machine learning models—Decision Tree, Support Vector Machine (SVM), Logistic Regression, and Random Forest—for our experiment. Each model was trained and tested on the original dataset, divided into 80\% for training and 20\% for testing. Figure \ref{fig-exp-without-obfuscated} summarizes the performance results. All models demonstrated high effectiveness, with SVM and Random Forest exhibiting the most consistent and superior performance. The Decision Tree model achieved an accuracy of 0.996 and an F1 score of 0.994, reflecting strong predictive capabilities. SVM outperformed all models, attaining an accuracy of 0.998 and an F1 score of 0.998, demonstrating exceptional detection ability. Logistic Regression also performed well, with 0.996 accuracy and an F1 score of 0.995. Random Forest closely matched SVM, achieving 0.998 accuracy and an F1 score of 0.998, further highlighting the robustness of ensemble-based methods in XSS detection.

\subsection{Evaluate Models with Obfuscated XSS}

In the following experiment, we created an obfuscated version of our XSS testing dataset by applying the obfuscation methods described in the approach section. Specifically, each original payload containing a malicious JavaScript vector was modified by randomly selecting one obfuscation technique (JavaScript obfuscation, Base64 encoding, URI encoding, or String-Splitting) with varying probabilities to simulate realistic adversarial variations. The resulting obfuscated payloads replaced the original payloads in our testing dataset. We then used this updated, obfuscated test dataset to evaluate machine learning models previously trained solely on non-obfuscated XSS data, thereby assessing their robustness and generalization capabilities against obfuscated attacks.

\begin{figure}[h]
  \centering
  \includegraphics[width=\linewidth]{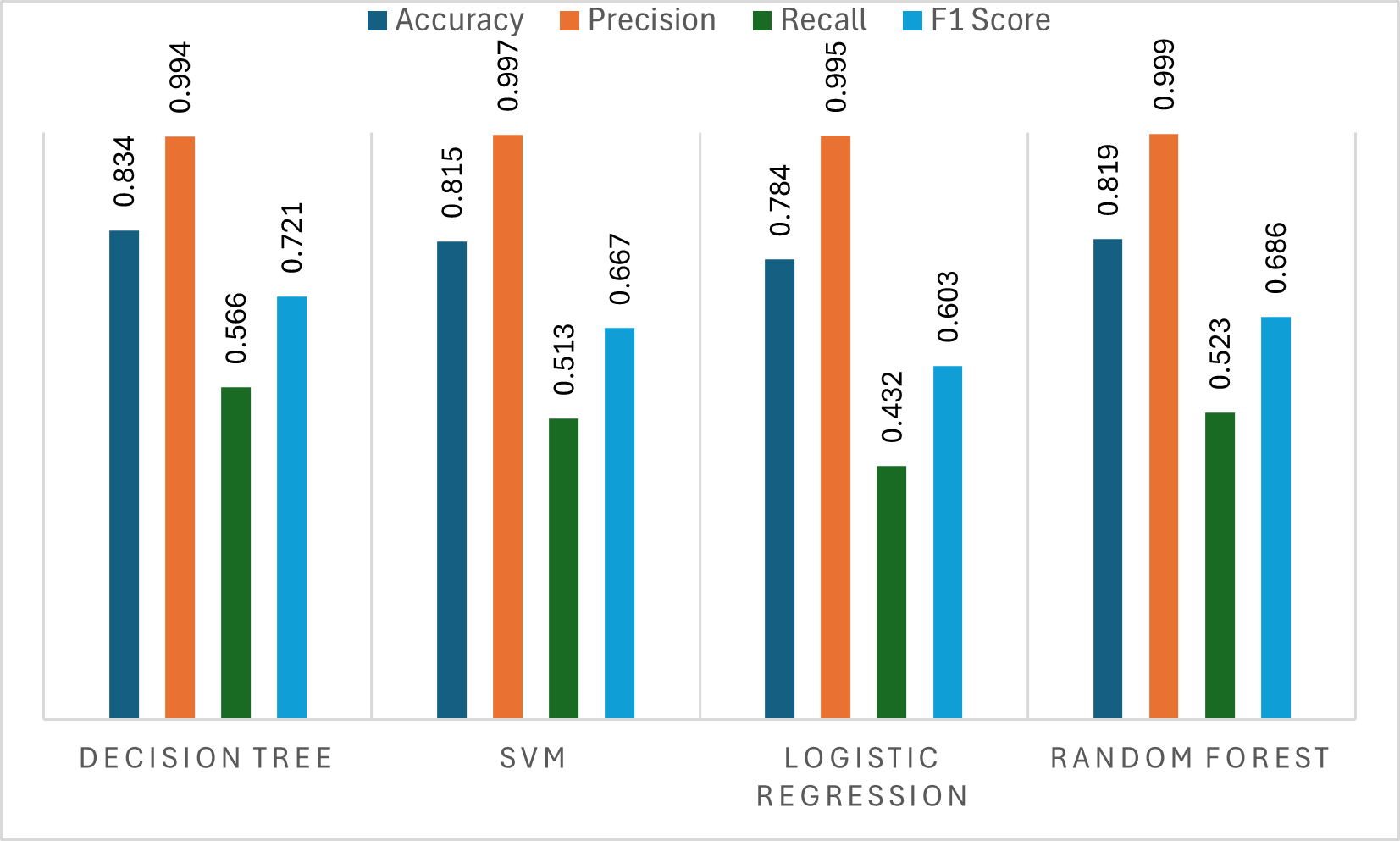}
  \caption{The performance of models when trained with original data and tested with obfuscated XSS}
  \label{fig-exp-with-obfuscated}
\end{figure}

The performance of all models significantly declined when tested on obfuscated XSS payloads, as illustrated in Figure \ref{fig-exp-with-obfuscated}. Accuracy dropped across all models, with SVM decreasing from 0.998 to 0.815 and Logistic Regression falling from 0.996 to 0.784. Recall was the most affected metric, demonstrating that models struggled to correctly identify obfuscated XSS payloads. Logistic Regression’s recall dropped from 0.992 to 0.432, and SVM’s recall declined from 0.997 to 0.513, indicating poor generalization to obfuscated samples. Despite the significant drop in accuracy and recall, precision remained high across all models, with Decision Tree maintaining 0.994, SVM at 0.997, Logistic Regression at 0.995, and Random Forest at 0.999. This suggests that when the models classified a sample as malicious, they were still mostly correct. However, their ability to detect obfuscated XSS payloads (low recall) was significantly weakened. Among the models, Random Forest and Decision Tree retained relatively better recall, with Random Forest dropping from 0.996 to 0.523 and Decision Tree from 0.992 to 0.566. This suggests that ensemble-based models may be more resilient to obfuscation techniques. The consistent drop in all performance metrics underscores the vulnerability of machine learning models when exposed to obfuscated data they were not trained on.

These findings emphasize the critical need to incorporate obfuscated XSS samples during training. Models trained solely on non-obfuscated payloads fail to generalize well to adversarially obfuscated inputs. By training on a mixture of original and obfuscated XSS data, machine learning models can improve their detection capabilities and robustness against real-world attacks that leverage obfuscation techniques.
 
\subsection{Leveraging LLM to Strengthen Models}

To enhance the detection capabilities of machine learning models, we fine-tuned a CodeT5-small model to generate obfuscated XSS payloads. This process aimed to produce obfuscated data that would diversify the training set and improve the robustness of ML models in identifying obfuscated attacks.  

% The training set comprised 80\% of the total data, while the remaining 20\% was allocated for testing. This division ensured that the models were evaluated on data they had not seen during training, providing a more accurate assessment of their generalization capabilities.

% \begin{table}[htbp]
% \caption{The performance of models after training LLM-generated Obfuscated XSS}
% \begin{center}
% \begin{tabular}{|l|c|c|c|c|}
% \hline

% \textbf{Model} & \textbf{Accuracy}& \textbf{Precision}& \textbf{Recall} & \textbf{F1 Score}  \\
% \hline
% Decision Tree& 0.950 & 0.997 & 0.872 & 0.930\\
% \hline
% Support Vector& 0.967 & 0.999 & 0.913 & 0.954 \\
% Machine& &  &  &  \\
% \hline
% Logistic Regression& 0.963 & 0.998 & 0.904 & 0.948\\
% \hline
% Random Forest& 0.995 & 0.999 & 0.988 & 0.993\\
% \hline

% \end{tabular}
% \label{tab-exp-with-LLM-Ob}
% \end{center}
% \end{table}

\begin{table}[htbp]
  \caption{The performance of models after training on LLM-generated obfuscated XSS and tested with obfuscated data}
  \label{tab:exp-with-LLM-Ob}
  \begin{tabular}{lcccc}
    \toprule
    \textbf{Model} & \textbf{Accuracy} & \textbf{Precision} & \textbf{Recall} & \textbf{F1 Score} \\
    \midrule
    Decision Tree & 0.950 & 0.997 & 0.872 & 0.930 \\
    SVM & 0.967 & 0.999 & 0.913 & 0.954 \\
    Logistic & 0.963 & 0.998 & 0.904 & 0.948 \\
    Regression &  &  &  &  \\
    Random Forest & 0.995 & 0.999 & 0.988 & 0.993 \\
    \bottomrule
  \end{tabular}
\end{table}

\noindent We incorporated the original training data with the LLM-generated obfuscated XSS and trained the models, resulting in a substantial improvement in performance compared to previous experiments where models were trained without obfuscated data. The results, as presented in Table \ref{tab:exp-with-LLM-Ob}, demonstrate a significant enhancement in detection capabilities. The Random Forest model achieved the highest accuracy at 0.995 with an F1 score of 0.993, highlighting its robustness in handling obfuscated payloads. The Support Vector Machine (SVM) also performed exceptionally well, attaining an accuracy of 0.967 and an F1 score of 0.954, indicating its strong generalization to obfuscated XSS attacks. Similarly, Logistic Regression demonstrated solid results, with an accuracy of 0.963 and an F1 score of 0.948. The Decision Tree model, while slightly behind the ensemble-based methods, still maintained a high accuracy of 0.950 and an F1 score of 0.930. These findings emphasize the importance of integrating LLM-generated obfuscations into the training process to significantly enhance the resilience of detection models against real-world, sophisticated XSS attacks.

\subsection{Complexity of LLM-generated Obfuscated XSS}
To evaluate the complexity of the LLM-generated obfuscated XSS payloads, we employed Shannon entropy \cite{shannon1948} as a metric. This method provides a quantifiable measure of unpredictability within the obfuscated code, allowing us to compare the complexity of LLM-generated samples against baseline samples produced by traditional obfuscation tools. Calculating entropy for each payload highlighted the level of randomness, indicating the difficulty for conventional models in parsing the content. The LLM produced unique and complex obfuscations using a higher temperature setting (1.5). The analysis showed that LLM-generated samples were, on average, 28.1\% more complex, increasing their potential to evade detection. However, higher temperatures sometimes resulted in syntactically invalid code, which, despite deviations, contributed to training robust Random Forest models that were more effective in detecting obfuscated XSS attacks.

\subsection{Semantic Equivalence Between Original and Obfuscated XSS}

Comprehensive functional testing was conducted to evaluate whether LLM-generated obfuscated XSS code retained the same functionality as the original payloads. The goal was to confirm that obfuscation changed only the code's structure without altering its behavior. Testing involved running both original and LLM-generated obfuscated XSS payloads in controlled browser environments to observe their actions. The primary measure of semantic equivalence was whether both types of payloads triggered the same JavaScript alerts or actions. Results showed that many LLM-generated obfuscations maintained functional parity with their original versions, demonstrating the LLM's capability to modify code structure while preserving behavior, which is essential for security testing. Challenges included limitations of the 60M parameter CodeT5 model and generation constraints that sometimes produced invalid or non-functional code. The complex training data from tools like the JavaScript obfuscator contributed to these difficulties, occasionally hindering valid output generation. JavaScript obfuscators often produce highly randomized and context-sensitive patterns, which can be difficult for smaller language models like CodeT5 (60M parameters) to generalize effectively. However, during fine-tuning, the model showed learning convergence, indicating its ability to replicate training data patterns and generate structurally complex outputs. Although ensuring semantic equivalence for highly intricate samples was challenging, the generated code displayed significant complexity, supporting effective training for machine learning models to detect obfuscated XSS.
\section{Related Works}

Using Large Language Models (LLMs) in software engineering has led to significant progress in automated bug detection, repair, and vulnerability analysis. Recent studies have demonstrated various ways LLMs help address complex security issues. BUGFARM, for example, employs LLMs to generate complex, hard-to-detect bugs by modifying less-examined areas of code, pushing the limits of machine learning-based bug detection and repair systems \cite{ibrahimzada2023automated}. In the area of vulnerability generation, QED is a model-checking tool designed to create attack vectors, such as Cross-Site Scripting (XSS) and SQL injection, differing from our work, which focuses on using LLM-generated obfuscated XSS data to enhance ML detection models \cite{martin2008automatic}. LIBRO showcases how LLMs can automate test case generation from bug reports, aiding developers in reproducing bugs more efficiently \cite{kang2023large}. LLbezpeky examines the ability of LLMs to detect Android vulnerabilities through prompt engineering and retrieval-augmented generation, proving the versatility of LLMs in various security tasks \cite{mathews2024llbezpeky}. Llm4sa automates the inspection of static analysis warnings, addressing the challenge of high false-positive rates and reducing manual review workload \cite{wen2024automatically}. Additionally, the study on token-level bug localization and repair, using models like CodeT5 and CodeGen, demonstrates how LLMs can achieve precise bug fixing and enhance the repair process \cite{hossain2024deep}. Our work builds on these studies by fine-tuning LLMs to generate complex, obfuscated XSS code, strengthening ML model training for advanced XSS detection, and complementing the broader automated bug generation and repair field.
\section{Conclusion}

This study demonstrates the importance of incorporating obfuscated XSS data into machine learning models for effective XSS detection. Our approach of fine-tuning a CodeT5-small LLM to generate complex obfuscated code proved to be a significant step forward. We showed that models trained on diverse datasets, including LLM-generated obfuscations, exhibited marked improvements in detecting obfuscated XSS payloads compared to models trained only on traditional datasets. The experimental results confirmed that including obfuscated samples enhances the robustness and resilience of ML-based XSS detection systems against real-world, complex attacks.

In future work, we aim to enhance the quality of LLM-generated obfuscated code, focusing on maintaining semantic equivalence between the original code and the obfuscated output. Further refinement of the fine-tuning process and the generation of XSS code by LLMs will be a crucial area of exploration. One potential approach is to incorporate the semantics of the original code during the fine-tuning phase and while generating obfuscated code, allowing the LLM to produce obfuscations that are both complex and functionally consistent with the original.

 \bibliographystyle{ACM-Reference-Format}
 \bibliography{WiSec}

% %%
% %% If your work has an appendix, this is the place to put it.
% \appendix









\end{document}